\documentclass[aps,prd,showpacs]{revtex4}
\usepackage{graphicx,color}
\usepackage{latexsym}
\usepackage{amsmath}
\usepackage{amsfonts}
\usepackage{amssymb}
\usepackage[latin1]{inputenc}

\newcommand{\be}{\begin{equation}}
\newcommand{\ee}{\end{equation}}
\newcommand{\ben}{\begin{eqnarray}}
\newcommand{\een}{\end{eqnarray}}
\newcommand{\sech}{\rm sech}
\newcommand{\arcsinh}{\rm arcsinh}

\begin{document}
\title{First-order framework and domain-wall/brane-cosmology correspondence}
\author{D. Bazeia$^{a}$, F.A. Brito,$^{b}$ and F.G. Costa,$^{a}$}
\affiliation{$^a$Departamento de F\'\i sica, Universidade Federal da
Para\'\i ba, Caixa Postal 5008, 58051-970 Jo\~ao Pessoa Para\'\i ba, Brazil\\
$^b$Departamento de F\'\i sica, Universidade Federal de Campina
Grande, Caixa Postal 10071, 58109-970 Campina Grande Para\'\i ba,
Brazil}
\date{\today}
\begin{abstract}
We address the possibility of finding domain wall solutions from
cosmological solutions in brane cosmology. We find first-order
equations for corresponding cosmology/domain wall solutions induced
on 3-branes. The quadratic term of energy density in the induced
Friedmann equation plays a non-standard role and we discuss the way the
standard cosmological and domain wall models are recovered as the
brane tension becomes large and show how they can be described by
four-dimensional supergravity action in such a limit. Finally, we
show that gravity on the 3-brane is locally localized as one moves
away from the two-dimensional domain walls living on the brane.
\end{abstract}
\maketitle

\section{Introduction}

The evolution of a 3-brane universe in the Randall-Sundrun scenario has been recently considered
in the literature \cite{rs,lcd,defayet}. The Einstein equations for braneworld
cosmology admits a first integral that governs the cosmological
evolution on the 3-brane, without any mention to the bulk evolution
behavior. This first integral involves a modified Friedmann
equation, in which the energy density contributes with both linear and
quadratic terms. The trace of the five-dimensional spacetime is
revealed at high energy by the quadratic term, whereas the
four-dimensional standard cosmology is recovered in the low energy
limit.

In this paper we consider the cosmological evolution on the 3-brane
driven by a real scalar field. We are able to find first-order
equations satisfying the equations of motion, by making a suitable
choice of the scalar field potential written in terms of a
`superpotential' in a non-standard way, with the standard scalar
potential of a four-dimensional supergravity theory being recovered at
relatively low energy.

In order to make a correspondence between brane cosmology solutions
and domain wall solutions living on the 3-brane,  we consider the
domain wall counterpart of the brane cosmology set up aforementioned,
by carrying out analytic continuations which leads the time
coordinate into a space coordinate on the 3-brane. At late times, the
familiar scalar potential of a four-dimensional supergravity is
recovered and then supergravity domain walls correspond to standard
cosmology. In this regime the correspondence here falls into the
framework already developed in Refs.~\cite{mcvetic95,cvetic97,bglm,abl,st2006.1,st2006.2}.

Concerning the global behavior of the analytic continued 3-brane
solution, our results show that gravity is localized on the 3-brane
with the most concentration where there exists two-dimensional
domain walls living in. As one goes far from such domain walls the
gravity tends to be locally localized on the brane, i.e., after
its falloff around the brane, the brane warp factor develops
returning points and goes back to infinity just as in the
Karch-Randall scenario \cite{kr}. On the domain walls, the brane is
flat assuming a four-dimensional Minkowski geometry \cite{rs},
whereas far from the domain walls the brane is bent \cite{abl,st2006.1}
and assumes a four-dimensional Anti-de-Sitter ($AdS_4$)
geometry -- see, e.g., Refs.~\cite{kalo,st,fwolf,nunez,alesio,bbg2004,
bbl2006,celi,dalla,cpv,ber,vau}.

The paper is organized as follows. In Sec.~\ref{1st} we present the
first-order framework for non-standard brane cosmology and consider
explicit examples. In Sec.~\ref{DW_Cosm}, we extend this framework
for domain walls solutions by carrying out an analytic
continuation and there we also consider some explicit examples. In
Sec.~\ref{warpSOL}, we show how the localization of gravity is
affected by the localization of the two-dimensional domain walls on
the brane. Finally in Sec.~\ref{conclu} we present our conclusions.

\section{First-order framework}
\label{1st}

In this paper we extend the first-order formalism
recently considered in \cite{bglm} to the case of the non-standard Friedmann
equation which appears in brane cosmology \cite{lidsay,ellis}. The metric describing the
cosmological evolution on a 3-brane is
\ben\label{metric_master}
ds_5^2=-n^2(t,r)dt^2+a^2(t,r)\gamma_{ij}dx^idx^j+b^2(t,r)dr^2,
\een
where $\gamma^{ij}$ is a maximally symmetric 3-dimensional metric
with spatial curvature $k=-1,0,1$. The five-dimensional Einstein
equations are found for the action of a 3-brane embedded in a
five-dimensional bulk, i.e.,
\ben\label{action}
S_5\!=\!-\frac{1}{2\kappa_{5}^2}\!\!\int{\!d^5x\sqrt{-g}(R+\Lambda_{bulk})
+\int{\!d^5x\sqrt{-g}\,\delta(r)}{\cal L}},
\een
where ${\cal L}$ describe the dynamics on the brane. Below we shall assume
$\kappa_4^2=8\pi G\simeq{\kappa_5^4\sigma}/{6}=2$. We are
interested in the case where the five-dimensional bulk is an $AdS_5$
spacetime \cite{defayet}, with the cosmological constant defined as
$\rho_{bulk}\equiv \Lambda_{bulk}=-\sigma^2\kappa_5^2/6$ which
satisfies the Randall-Sundrum fine tuning \cite{rs}.

Let us start with the induced Friedmann equation on the brane given by
\ben\label{friedman}
H^2=\frac{2}{3}\rho\left(1+\frac{\rho}{2\sigma}\right),
\een
where $H={\dot{a}_0}/{a_0}$ and $a_0\!=\!a(t,r\!=\!0)$ is 
the scale factor on the brane worldvolume with the metric
\ben\label{metric4d}
ds_4^2=-dt^2+ a_0^2(t)(dx^2+dy^2+dz^2).
\een
Here $\rho$ is the energy density on the brane, $\sigma$ is the brane tension and
$\rho_{bulk}\neq0$ was recast in terms of $\sigma$. The quadratic
nature of the density $\rho$ is due to junction condition across the
3-brane \cite{defayet} embedded in the bulk with five dimensions.
The equation involving the pressure $p$ is
\ben\label{aceler}
\dot{H}=-(\rho+p)\left(1+\frac{\rho}{\sigma}\right).
\een
The equations (\ref{friedman}) and (\ref{aceler}) can be found from the
Einstein equations on the 3-brane \cite{maeda},
$^{(4)}G_{\mu\nu}=-\Lambda_4q_{\mu\nu}+\kappa^2_{(4)}T_{\mu\nu}+
\kappa^4_{(5)}\pi_{\mu\nu}-E_{\mu\nu}$, where $\Lambda_4$ and
$q_{\mu\nu}$ are the cosmological constant and the metric on the
brane, respectively. $\pi_{\mu\nu}$ is quadratic in the
energy-momentum tensor $T_{\mu\nu}$ and $E_{\mu\nu}$ is part of the
five-dimensional Weyl tensor. The standard cosmology is recovered as
$\sigma$ becomes sufficiently large, i.e.,
$\kappa_{(5)}^4\sim1/\sigma\sim0$. Since, in our case, we are
disregarding $\Lambda_4$ and $E_{\mu\nu}$, the brane dynamics in this
regime is governed by the effective action
\ben\label{4dA}
S^{\rm eff}_4=-\frac{1}{2\kappa^2_{(4)}}\int{d^4x\sqrt{-q}\,\left(\,
^{(4)}R-2\kappa^2_{(4)}\cal{L}\,\right)}.
\een
This equation will be useful for identifying a four-dimensional `supergravity' action later. Let us
now assume that the brane cosmology is driven by a scalar field whose
Lagrangian density is
\ben\label{4dL}
{\cal L}=-\frac{1}{2}\partial_\mu\phi\partial^\mu\phi-V(\phi),
\een
with $\mu=0,1,2,3.$ Thus, the familiar equations for the energy density
$\rho$ and the pressure $p$ are
\ben\label{rho}
\rho=\frac{1}{2}\dot{\phi}^2+V(\phi),\\
\label{p} p=\frac{1}{2}\dot{\phi}^2-V(\phi). 
\een
By applying the induced equation of conservation for the energy density on the
brane, i.e., $\dot{\rho}+3H(\rho+p)=0$, we find \ben\label{rho_dot}
\dot{\rho}=-3H\dot{\phi}^2. \een The scalar field dynamics is
governed by the equation of motion
\ben\label{eomPhi}
\ddot{\phi}+3H\dot{\phi}+\frac{\partial V}{\partial\phi}=0.
\een
Since $\rho\equiv\rho(\phi)$, we use the fact that
$\dot{\rho}=\rho'(\phi)\dot{\phi}$, such that Eq.~(\ref{rho_dot}) becomes
\ben\label{phi_dot1}
\dot{\phi}=-\frac{\rho'(\phi)}{3H}.
\een

To get to the first-order equations, we follow the procedure of \cite{bglm}, e.g. we introduce $W=W(\phi)$
and define $H={\dot{a}_0}/{a_0}=W(\phi)$, such that we have the following first-order equation
\ben\label{warp}
\frac{\dot{a}_0}{a_0}=W(\phi),
\een
which allows us to rewrite the Eq.~(\ref{friedman}) in the form
\ben\label{friedman2}
\rho^2+2\sigma\rho-3\sigma W^2=0.
\een
This algebraic equation has the following solutions
\ben\label{rho_sols}
\rho_\pm={-\sigma\pm\sigma\sqrt{1+\frac{3W^2}{\sigma}}}.
\een
We consider the solution with the upper sign, because of the
positive energy condition. Thus, by differentiating the
Eq.~(\ref{rho_sols}) it is not difficult to find that
\ben\label{rho_phi_p}
\rho'(\phi)=\frac{3WW_\phi}{\sqrt{1+\frac{3W^2}{\sigma}}}.
\een
The above Eq.~(\ref{phi_dot1}) can be now written as a first-order equation
for the scalar field $\phi$ and the `superpotential' $W$, i.e.,
\ben\label{phi_dot2}
\dot{\phi}=-\frac{W_\phi}{\sqrt{1+\frac{3W^2}{\sigma}}}.
\een
One can easily check that the two first-order equations
(\ref{warp}) and (\ref{phi_dot2}) satisfy the two second-order
equations (\ref{aceler}) and (\ref{eomPhi}).

The scalar potential $V(\phi)$ can be found via Eqs.~(\ref{rho}),
(\ref{rho_sols}) (upper sign), and (\ref{phi_dot2}). It has the explicit form
\ben\label{V_phi}
V(\phi)=-\sigma+
\sigma\sqrt{1+\frac{3W^2}{\sigma}}-\frac{1}{2}\frac{W_\phi^2}
{{1+\frac{3W^2}{\sigma}}}.
\een
For a large brane tension $\sigma$,
such that $W^2/\sigma\ll1$, one can expand the potential in  a power series as
\ben\label{V_phi_series}
V(\phi)&=&\frac{3}{2}W^2-\frac{1}{2}W_\phi^2
+\frac{3}{2}W_\phi^2\frac{W^2}{\sigma}+\cdots
\een
where the standard potential \cite{bglm} is recovered by taking into account only
the familiar {\it quadratic} terms
\ben\label{Vsugra}
V(\phi)&\simeq&\frac{3}{2}W^2-\frac{1}{2}W_\phi^2.
\een
Under similar approximations the standard first-order equation is recovered
by turning on only {\it linear} terms in Eq.~(\ref{phi_dot2}).

On the other hand, for small brane tension, such that
$W^2/\sigma\gg1$, the scalar potential approaches
\ben\label{V_phi_leading}
V(\phi)&\simeq&\sigma\sqrt{\frac{3}{\sigma}}|W|,
\een
and the first-order equation now reads
\ben\label{phi_dot3}
\dot{\phi}\simeq-\sqrt{\frac{\sigma}{3}}\frac{W_\phi}{|W|}.
\een
Substituting (\ref{V_phi_leading}) and (\ref{phi_dot3}) into the
second-order equation (\ref{eomPhi}), one finds that
$\ddot{\phi}\simeq0$, for $W>0$. This is precisely the slow-roll
regime. Since $\ddot{\phi}\simeq0$, thus for consistency
$\dot{\phi}\simeq const.$ This implies that we can determine the
`superpotential', the inflaton solution for (\ref{phi_dot3}), and
the scale factor solution for (\ref{warp}) given by
\ben\label{Wf}W(\phi)=V_0\,e^{\alpha\phi}, \qquad
\phi(t)=-\sqrt{\frac{\sigma}{3}}\,\alpha(t-t_0), \qquad a_0(t)=
a_0\exp{\left[\frac{-V_0\sqrt{3}}{\alpha^2\sqrt{\sigma}}
\exp{\left(\frac{-\alpha^2\sqrt{\sigma}}{\sqrt{3}}(t-t_0)\right)}\right]},
\een
where $\alpha$ and $V_0$ are constants, being
$V_0^2/\sigma\gg1$ consistent with the regime $W^2/\sigma\gg1$. For
$\alpha>0\,\, (\alpha<0)$, the Universe develops inflation at later
(earlier) times $t<t_0$. The exponential potential is also
consistent with string/M-theory.

\begin{figure}[ht]
\centerline{\includegraphics[{width=7.0cm}]{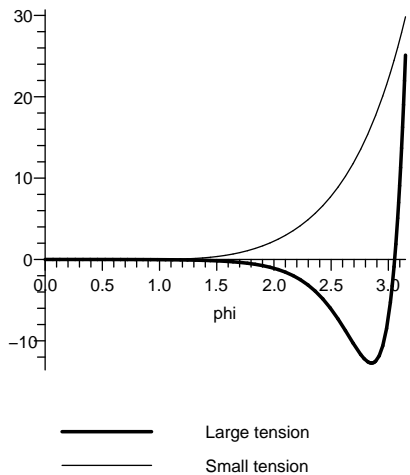}}
\caption{The scalar potential $V(\phi)$ for small
($\sigma=1$) and large ($\sigma=3\times 10^3$) tension, with
$V_0=1/20$ and $\alpha=\sqrt{7/2}$. For large $\sigma$, the minimum
occurs at $\phi_0=2.85$ such that our Universe (infinite and flat)
can undergo an oscillatory expansion.}\label{fig0}
\end{figure}

Conversely, for $V_0^2/\sigma\ll1$ one gets to the
regime $W^2/\sigma\ll1$, and the potential (\ref{V_phi_leading}) and the
first-order equation (\ref{phi_dot3}) do not make sense anymore.
Instead, in this regime one recovers the previous analysis with the scalar
potential (\ref{Vsugra}). Thus  $V_0^2/\sigma$ is assumed to be the
coupling connecting asymptotic regimes of the exact potential
(\ref{V_phi}) given in the form
\ben\label{V_phi_exp}
V(\phi)=-\sigma+\sigma\sqrt{1+\frac{3V_0^2}{\sigma}\,e^{2\alpha\phi}}-
\frac{1}{2}\frac{V_0^2\alpha^2e^{2\alpha\phi}}
{{1+\frac{3V_0^2}{\sigma}\,e^{2\alpha\phi}}}.
\een
As $\alpha>\sqrt{3}$, this potential develops a minimum at
\ben\phi_0=\frac{1}{\alpha}\ln{\left[\frac{\sqrt{\alpha^{4/3}3^{1/3}\sigma-3\sigma}}{3V_0}\right]}.
\een
To ease comparison, in Fig.~\ref{fig0} we depict the two regimes.
Note that as $V_0^2/\sigma$ increases, the scalar potential
approaches the form given in (\ref{V_phi_leading}. On the other hand,
as $V_0^2/\sigma$ decreases, the scalar
potential (\ref{V_phi_exp}) approaches the form (\ref{Vsugra}). In
this regime, the rolling inflaton field could eventually achieve the
vacuum
\ben \Lambda\equiv V(\phi_0)=-\frac{1}{2} e^{2\alpha \phi_0}
(-3+\alpha^2) V_0^2.
\een
It is clear from the equation above, that
for $\alpha>\sqrt{3}$, one finds $\Lambda<0$, which corresponds to
an $AdS_4$ vacuum. Moreover the negative vacuum at $\phi_0$
signalizes the possibility of the Universe undergoing an oscillatory
expansion \cite{ts}--- see Fig.~\ref{fig0}. Recall that, in our
model, the Universe is infinite and flat $(k=0)$. For
$\alpha<\sqrt{3}$, the potential has only the minimum $\phi_0=0$,
and $\Lambda=-(1/2)(-3+\alpha^2)V_0^2>0$ corresponds to a $dS_4$
vacuum. This produces inflation only at late times.

The cosmological scenario discussed above is particularly
interesting, because the coupling $V_0^2/\sigma$ can vary as the
brane inflates, such that the potentials pictured in Fig.~\ref{fig0}
are asymptotic limits of a same scalar potential governing the brane
evolution at high and low energy scales. However, other interesting
cosmological scenarios are those where one fixes $V_0^2/\sigma=1$,
which allows for a wider class of `superpotentials'. In the
following we shall investigate such scenarios by using examples
where $W\propto\sqrt{\sigma}$ all times.

\subsection{Cosmological Examples}

\begin{figure}[ht]
\centerline{\includegraphics[{angle=90,height=7.0cm,angle=180,width=8.0cm}]{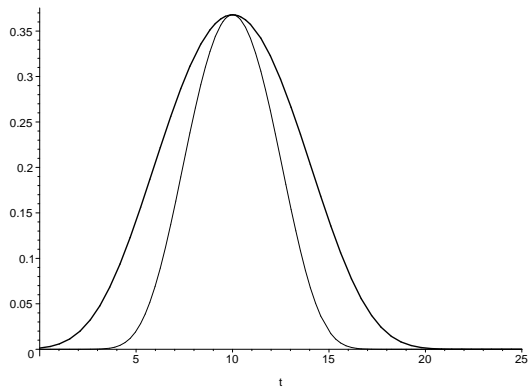}}
\caption{The scale factor $a_0(t)$ for $\sigma=1/2$ (thin
line) and for $\sigma=1/5$ (thick line). The larger (smaller)
 $\sigma$ favors later (earlier)  inflation. Note the end of inflation at $t_0=10$.}\label{fig1}
\end{figure}

Given that many examples \cite{bglm,abl,st2006.1,st2006.2} have been
previously considered in the literature for low energy limit of
theory we describe above, let us now consider some examples for the
exact theory. For $W(\phi)=\sqrt{\sigma/3}\sinh{\phi}$, the
Eq.~(\ref{phi_dot2}) is easily integrated whose solution is simply
\ben\label{phi_sol_W1} \phi=-\sqrt{\frac{\sigma}{3}}(t-t_0). \een
The scale factor $a_0(t)$ on the brane can also be readily found by
using the equation (\ref{warp}) and the solution (\ref{phi_sol_W1}).
Its form is given by
\ben\label{a_scale}a_0=\exp{\left[-\cosh{\sqrt{\frac{\sigma}{3}}}(t-t_0)\right]}.
\een
The larger is the brane tension (i.e., the standard cosmology
regime), the later the inflation occurs. On the other hand, for small
brane tension one deviates from the standard cosmology and inflation
occurs only at earlier times --- See Fig.~\ref{fig1}. Note that the
end of inflation occurs at a time $t_0$, with decelerating
universe for $t>t_0$, in a way similar to the case of quadratic chaotic
inflation models \cite{linde}.

\begin{figure}[ht]
\centerline{\includegraphics[{angle=90,height=7.0cm,angle=180,width=8.0cm}]{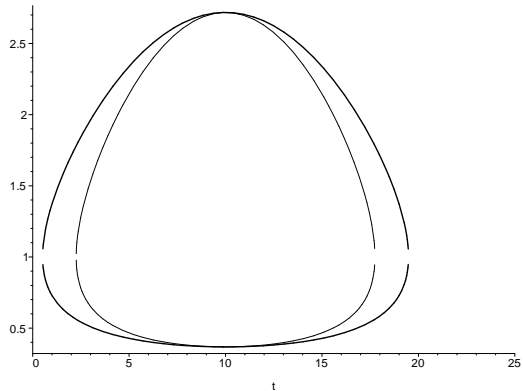}}
\caption{The scale factor $a_{0\pm}(t)$ for $\sigma=1/20$ (thin
line) and for $\sigma=1/30$ (thick line). Note the two possibility of limited
expansion, with inflation beginning or ending at
$t_0=10$.}\label{fig2t}
\end{figure}

Let us now consider the example with $W(\phi)=\sqrt{\sigma/3}\tan{(\phi)}$. Here the
solutions are given by
\ben\label{ex2_phit}\phi=\mp\arcsin{\Big(\sqrt{\frac{\sigma}{3}}
\mbox{$(t-t_0)$}\Big)}
\een
and
\ben
{a_0}_\pm=\exp{\left[\pm\sqrt{-\frac{\sigma}{3}(t-t_0)^2+1}\,\,\right]}\label{ex2_at}.
\een
The inflaton field (\ref{ex2_phit}) behaves in a singular way.
The scale factor is depicted in Fig.~\ref{fig2t}. Note the two
possibility of `limited' expansion $a_+$ and $a_-$, with inflation
beginning ($a_{0-}$) or ending ($a_{0+}$) at $t_0=10$.

\section{Domain-Wall/Brane-Cosmology Correspondence}
\label{DW_Cosm}

It is now well-known that one can use cosmological solutions to find
domain wall solutions, and vice-versa, by making use of analytic
continuation \cite{cvetic93,cvetic97,ssakura2002,ssakura2002.1,st2006.1,st2006.2}.

All the developments above can be extended to give rise to domain wall
solutions living on the four-dimensional brane world-volume. To
carry out analytic continuation we make
\ben
&&W\to i\widetilde{W},\\ &&H\to i\widetilde{H},\\ &&t\to iy, \\
&&y\to it,
\een
where $\widetilde{H}={a_0'}/{a_0}=-\widetilde{W}.$ The original
four-dimensional metric (\ref{metric4d}) describing cosmological
solutions on the brane can now be written as
\ben\!
ds_4^2=dy^2+{a_0}^2(y)(-dt^2+dx^2+dz^2).
\een
This metric represents solutions of two-dimensional flat domain walls within asymptotically
four-dimensional Minkowski ($\mathbb{M}_4$) or anti-de Sitter
(${AdS}_4$) spacetime \cite{cvetic93,cvetic97}. These domain walls
are of current interest to cosmology \cite{spe,co}.

Since the domain wall solutions are analytic continued from the previous cosmological
solution, we cannot find asymptotically four-dimensional de Sitter
(${dS}_4$) spacetime here. Now, the first-order equations are given by
\ben\label{phi_prime}
&&\phi'=\frac{\widetilde{W}_\phi}{\sqrt{1-\frac{3\widetilde{W}^2}{\sigma}}},\\
\label{A_prime}&&\frac{a_0'}{a_0}=-\widetilde{W}.
\een
These first-order equations satisfy the second-order equations
(\ref{aceler}) and (\ref{eomPhi}) by properly carrying out the
analytic continuation. The scalar potential assumes the form
\ben\label{V_phi_tilde}
\widetilde{V}(\phi)=-\sigma+\sigma\sqrt{1-\frac{3\widetilde{W}^2}{\sigma}}+\frac{1}{2}\frac{\widetilde{W}_\phi^2}
{{1-\frac{3\widetilde{W}^2}{\sigma}}}.
\een
As in the previous case,
in the limit $\widetilde{W}^2/\sigma\ll1$ we get
\ben\label{Vsugra2}
\widetilde{V}(\phi)\simeq\frac{1}{2}\widetilde{W}_\phi^2-\frac{3}{2}\widetilde{W}^2.
\een
The Eqs.~(\ref{4dA})-(\ref{4dL}) and (\ref{Vsugra2}) can be
identified with the bosonic sector of a four-dimensional
supergravity theory \cite{mcvetic95,cvetic97,st2006.1,st2006.2}. Some important
comments are in order. The superpotentials are clearly connected as
$W^2\longleftrightarrow-\widetilde{W}^2$
$W_\phi^2\longleftrightarrow-\widetilde{W}_\phi^2$. We note that in
the limit of low energy ($\widetilde{W}^2/\sigma\ll1$ or
${W}^2/\sigma\ll1$) the potentials (\ref{Vsugra}) and
(\ref{Vsugra2}) are related as
${V}(\phi)\longleftrightarrow-\widetilde{V}(\phi)$ in the brane,
although this is not the case for the exact potentials, as we can
see from Eqs.~(\ref{V_phi}) and (\ref{V_phi_tilde}). This
identification would be possible if we also made $\sigma\to
-\sigma$, which would require another brane, together with the
analytic continuation. In doing so, the domain wall/brane cosmology
correspondence would be possible only between branes with tension of
reversed signals. Since the exact potentials hold at both high and
low energy regimes, let us consider the following reasoning: at high
energies, a large amount of branes $(\sigma)$ and anti-branes
($-\sigma$) is favored, such that the correspondence in a
brane-anti-brane pair $(-\sigma,\sigma)$ takes place for exact
potential ${V}(\phi)$ at one brane and $-\widetilde{V}(\phi)$ at the
other. Because branes and anti-branes tend to annihilate, at low
energy regime an asymmetry in the brane-anti-brane number ends up
favoring the correspondence in a single brane $\sigma$ (or
$-\sigma$). This is precisely what the non-exact potential
identification above in the limit of low energy indicates.

For domain walls, the superpotential should be a limited function,
i.e., $|\widetilde{W}|\leq\sqrt{\sigma/3}$. Thus, at the vacua
$\widetilde{W}_\phi=0$, the potential can only reach the values
$\widetilde{V}(\phi_{\rm vac})=-\sigma$ (${AdS}_4$ spacetime) or
$\widetilde{V}(\phi_{\rm vac})=0$ ($\mathbb{M}_4$ spacetime), as we
have already anticipated. Such a restriction on $\widetilde{W}$
helps us to choose an acceptable superpotential in a smaller set of
functions. Many limited functions we can investigate though our
preferred examples here will be those that can be integrated
analytically. Functions such as $\cos(\phi)$, $\sin(\phi)$,
$\tanh(\phi)$ and $\sech(\phi)$ are good examples. The domain-wall/brane-cosmology
correspondence, with the restricted set of
superpotentials for domain wall solutions, can guide ourselves to
find corresponding cosmological solutions in a smaller set of
`superpotentials'.

\subsection{Domain Wall Examples}

\begin{figure}[ht]
\centerline{\includegraphics[{angle=90,height=7.0cm,angle=180,width=8.0cm}]{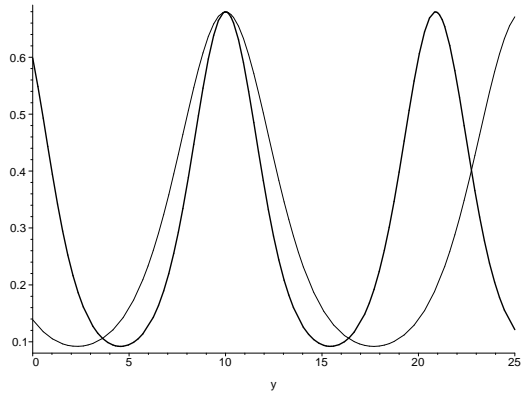}}
\caption{The ``warp'' factor $a_0(y)$ for $\sigma=1/2$ (thin
line) and for $\sigma=1$ (thick line).}\label{fig2}
\end{figure}

Let us consider some examples. One of them is the
analytic continued example that we obtain from the cosmological one:
$\widetilde{W}(\phi)=\sqrt{{\sigma}/{3}}\sin\phi$. The first-order
equations (\ref{phi_prime}) and (\ref{A_prime}) can be easily
integrated to give the simple solution \ben
\phi=\sqrt{\frac{\sigma}{3}}(y-y_0),\een and
\ben\label{a_scale_prime}a_0=\exp{\left[\cos{\sqrt{\frac{\sigma}{3}}}(y-y_0)\right]}.
\een This solutions is depicted in Fig.~\ref{fig2}. Note that it
represents an array of domain walls, centered around $y_0\!=\!10,
20,30,\dots$

\begin{figure}[ht]
\centerline{\includegraphics[{angle=90,height=7.0cm,angle=180,width=8.0cm}]{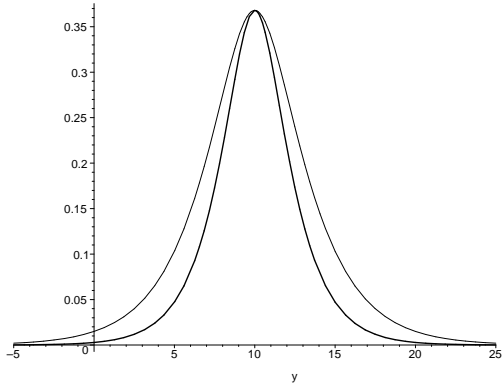}}
\caption{The `warp' factor $a_{0-}(y)$ for $\sigma=1/2$ (thin
line) and for $\sigma=1$ (thick line), centered around
$y_0=10.$}\label{fig3}
\end{figure}

We now consider the example
$\widetilde{W}(\phi)=\sqrt{{\sigma}/{3}}\tanh(\phi)$. The solutions are
\ben\label{ex2_phiy}\phi=\pm\arcsinh{\Big(\sqrt{\frac{\sigma}{3}}
\mbox{$(y-y_0)$}\Big)},
\een
and
\ben
{a_0}_{\mp}=\exp{\left[\mp\sqrt{\frac{\sigma}{3}(y-y_0)^2+1}\,\,\right]}\label{ex2_ay}.
\een
The kink--anti-kink profile which appears from (\ref{ex2_phiy}) connect the same
vacua. In spite of this, the geometry (\ref{ex2_ay}) has totally
different asymptotic behavior, i.e., whereas the `warp' factor $a_{0+}$
diverges the `warp' factor $a_{0-}$ does not. The non-divergent `warp'
factor $a_{0-}$ is depicted in Fig.~\ref{fig3}.

\begin{figure}[ht]
\centerline{\includegraphics[{angle=-90,width=7.0cm}]{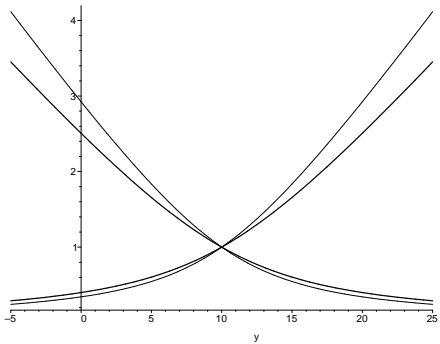}}
\caption{The ``warp'' factor $a_{0\mp}(y)$ for $\sigma=1/20$
(thin line) and for $\sigma=1/30$ (thick line), centered around
$y_0=10.$ For $y\to\infty$ ($y\to-\infty$), $a_{0+}$ ($a_{0-}$)
diverges.}\label{fig4}
\end{figure}

Another interesting example is given by
$\widetilde{W}(\phi)=\sqrt{{\sigma}/{3}}\,\sech(\phi)$. The solutions are
\ben\label{ex3_phiy}
\phi=\mp\arcsinh{\Big(\sqrt{\frac{\sigma}{3}}
\mbox{$(y-y_0)$}\Big)},
\een
and
\ben
{a_0}_{\mp}=\left[\sqrt{\frac{\sigma}{3}}(y-y_0)+\sqrt{\frac{\sigma}{3}(y-y_0)^2+1}\,\,\right]^{\mp1}
\label{ex3_ay}.
\een
Again the kink--anti-kink profile which appears from (\ref{ex3_phiy}) connect the same vacua. However, the corresponding
geometry (\ref{ex3_ay}) diverges asymptotically, i.e., for
$y\to\infty$ ($y\to-\infty$), $a_+$ ($a_-$) diverges. The solutions
$a_{0+}$ and $a_{0-}$ can be patched together at $y_0=10$ to form a well
behaved ``warped'' geometry for the domain walls on the brane. The
solutions are pictured in Fig.~\ref{fig4}.

\section{The global behavior of the brane solution}
\label{warpSOL}

Here we will examine how the analytic continued warp factor $a(t\to
iy,r)$ feels the effect of the domain wall solutions on the brane.
We specially investigate the behavior of the warp factor as we move
far from the domain wall, such that the brane geometry changes from
a Minkowski ($\mathbb{M}_4$) to an asymptotically $AdS_4$ geometry.

The original time-dependent warp factor solution for a 3-brane
embedded in $AdS_5$ spacetime \cite{defayet} is given by
\ben\label{sol_axr} a(t,r)\!=\!
\left[\left(1+\frac{\kappa_5^2\rho_b^2}{6\rho_{bulk}}\right)\frac{a_0^2}{2}
+\left(1-\frac{\kappa_5^2\rho_b^2}{6\rho_{bulk}}\right)\frac{a_0^2}{2}\cosh(\mu
r)\right.\left.-\frac{\kappa_5\rho_b}{\sqrt{-6\rho_{bulk}}}a_0^2\sinh(\mu
|r|)\right]^{1/2},\!\!
\een
where $\mu=(1/3)(\kappa_5^4\sigma^2)^{1/2}$. We have disregarded the
radiation `${\cal C}-$term' and the curvature `$k-$term' and applied
the Randall-Sundrum fine tuning. We thus recast
$\rho_{bulk}=-\sigma^2\kappa_5^2/6$ and recognize the energy density
on the brane as
\ben
\rho_b=\rho+\sigma=\sigma\sqrt{1-\frac{3\widetilde{W}^2}{\sigma}}.
\een
By carrying out the analytic continuation previously discussed,
the scale factor and brane energy density in (\ref{sol_axr}) changes
as $a_0(t)\to a_0(y)$ and $\rho_b(t)\to\rho_b(y)$. Recalling that
$\widetilde{W}=-{a_0'}/{a_0},$ we write the metric solution
(\ref{sol_axr}) in terms of the domain wall `warp' factor $a_0(y)$
as \ben\label{a_y_r}
a(y,r)=\left[\frac{3}{2}\frac{{a_0'}^2(y)}{a_0^2(y)\sigma}
+\left(1-\frac{3}{2}\frac{{a_0'}^2(y)}{a_0^2(y)\sigma}\right)\cosh(\mu
r)\right.-\left.\sqrt{1-3\frac{{a_0'}^2(y)}{a_0^2(y)\sigma}}\sinh(\mu|r|)\right]^{1/2}a_0(y).
\een

\begin{figure}[ht]
\centerline{\includegraphics[{angle=-90,width=7.0cm}]{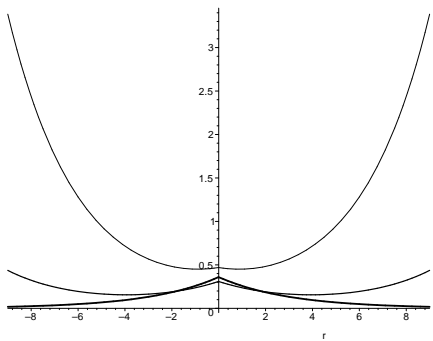}}
\caption{The warp factor $a(y,r)$ at $y=y_0$ (thick line),
$y=y_0+1$ (thin line), and $y=y_0+5$ (thinner line), being $y_0=10,$
 $\sigma=1$, and $\kappa_5^2=2$.}\label{fig5}
\end{figure}

The global behavior of the metric (\ref{sol_axr}) depending on the
domain walls living inside the brane is depicted in Fig.~\ref{fig5}.
The figure shows the localization of gravity on the brane changing
as we move away from the domain wall --- here we applied the
solution (\ref{ex2_ay}). The brane warp factor is peaked around the
brane centered at $r\!=\!0$ as we are settled on the domain walls
centered at $y=y_0$. However, the brane warp factor presents
returning points \cite{kr} as we move far from the domain walls at
the positions, say, $y=y_0+1$, $y=y_0+5$, and so on.
This behavior shows that at $y=y_0$ the geometry on the brane approaches a
Minkowski ($\mathbb{M}_4$) geometry which leads to a global
localization of gravity on the brane \cite{kr,bbg2004}.

On the other hand,
the more we move to positions far from the domain walls the more we
approach a vacuum with negative cosmological constant
$\widetilde{V}(\phi_{\rm vac})=-\sigma$ inside the brane which
implies an $AdS_4$ geometry on the brane. In such regime the brane
warp factor after falling off tends to turn around and grow toward
infinity. This reproduces the effect of `locally localized gravity'
encountered in the Karch-Randall scenario \cite{kr,bbg2004} --- see e.g.
\cite{lykken} for more recent investigations. The main point here is
that now one can understand the changing of the `cosmological
constant' on the brane through the presence of domain walls.

\section{Conclusions}
\label{conclu}

In this paper we have addressed the issue of otbaining first-order equations
and making a correspondence between {\it brane cosmology solutions} to domain wall solutions, by
carrying out analytical continuation of the brane cosmology solution
\cite{defayet}.

We have been able to find first-order equations that satisfy the second
order equations governing the geometry and the scalar field on the brane.
We have shown that at the low energy limit, the first-order equations can be
related to the same first-order equations found in four-dimensional supergravity
action \cite{mcvetic95,cvetic97}. In this limit one recovers a correspondence
similar to the {\it domain-wall/cosmology} correspondence, well
discussed recently in the literature \cite{st2006.1,st2006.2}, where
the first-order equations for domain walls associated with Killing
spinors equations are identified with the first-order equations for
cosmology. An important point should be noted here. In the usual
domain-wall/cosmology correspondence, a $(d-1)$-brane solution,
regarded as a $(d-1)$-dimensional domain wall solution, is
analytically continued to play the role of a cosmological solution
in a $(d+1)$-dimensional FRW spacetime, and vice-versa. Differently,
in the present work only the domain-wall and cosmological brane solutions inside
the 3-brane are elements involved in the correspondence. However, it happens
that the {\it domain-wall/brane-cosmology} correspondence
on the 3-brane is similar to the usual domain-wall/cosmology in the
low energy limit. At this regime a supergravity at four-dimensions that comes out on the brane,
essentially carries most of the characteristics of the
$(d-1)$-dimensional domain-wall/cosmology correspondence \cite{mcvetic95,cvetic97,st2006.1,st2006.2}.
Several issues are still open to be addressed elsewhere, such as
investigating the correspondence pointed out here in a higher
dimensional domain-wall/brane-cosmology correspondence in low and
high energy limits.

As a consequence of the correspondence, we have found another interesting result,
which shows that the corresponding domain wall solutions play an
interesting role on the brane. At asymptotic limits, they connect
Minkowski ($\mathbb{M}_4$) geometry to $AdS_4$ geometry on the
3-brane. Thus, they are closely related with global and
local localization of gravity on the brane \cite{kr,bbg2004}. A point to
be naturally addressed in this new framework would be to investigate
the graviton spectrum by perturbing the analytically continued brane
cosmology solution.


The authors would like to thank CAPES, CNPq, and MCT-CNPq-FAPESQ for partial financial support.


\end{document}